\documentstyle{article}
\begin{document}

\baselineskip=23pt

\begin{flushleft}
{\bf {\huge Causal relation between regions I and IV of the Kruskal extension%
}}\\

\vspace{2mm}

{\bf Yi-Ping Qin$^{1,2,3}$ }

{\bf $^{1}$ Yunnan Observatory, Chinese Academy of Sciences, Kunming, Yunnan
650011, P. R. China}

{\bf $^{2}$ National Astronomical Observatories, Chinese Academy of Sciences 
}

{\bf $^{3}$ Chinese Academy of Science-Peking University joint Beijing
Astrophysical Center }
\end{flushleft}

\vspace{2mm}

\begin{center}
{\Large Summary}\\
\end{center}

By extending the exterior Schwarzschild spacetime in two opposite directions
with the Kruskal method, we get an extension which has the same $T-X$
spacetime diagram as has the conventional Kruskal extension, while allowing
its regions $I$ and $IV$ to correspond to different directions of the
original spacetime. We further extend the exterior Schwarzschild spacetime
in all directions and get a 4-dimensional form of the Kruskal extension. The
new form of extension includes the conventional one as a part of itself.
From the point of view of the 4-dimensional form, region $IV$ of the
conventional extension does not belong to another universe but is a portion
of the same exterior Schwarzschild spacetime that contains region $I$. The
two regions are causally related: particles can move from one to the other. %
\\\vspace{6mm}

It is known that the conventional Kruskal extension contains four parts of
spacetime, a black hole, a white hole, and two regions of the exterior
Schwarzschild spacetime [1]. The two regions were believed to belong to
different universes, and messages can not be exchanged between them. In the
following we show that this interpretation is not correct.

The standard metric form of the exterior Schwarzschild spacetime is [2] 
\begin{equation}
ds^2=-(1-{\frac{2M}r})dt^2+{\frac 1{1-{\frac{2M}r}}}dr^2+r^2(d\theta ^2+\sin
^2\theta d\phi ^2)\qquad \qquad (r>2M).
\end{equation}
Let 
\begin{equation}
\left\{ 
\begin{array}{c}
t=t \\ 
x=r\sin \theta \cos \phi  \\ 
y=r\sin \theta \sin \phi  \\ 
z=r\cos \theta 
\end{array}
\qquad \qquad (r>2M).\right. 
\end{equation}
Then (1) can be written as 
\begin{equation}
\begin{array}{c}
ds^2=-(1-{\frac{2M}r})dt^2+[\frac{2Mx^2}{r^2(r-2M)}+1]dx^2+[\frac{2My^2}{%
r^2(r-2M)}+1]dy^2+[\frac{2Mz^2}{r^2(r-2M)}+1]dz^2 \\ 
\\ 
+\frac{4M}{r^2(r-2M)}(xydxdy+yzdydz+zxdzdx)\qquad \qquad \qquad \qquad
(r>2M),
\end{array}
\end{equation}
where 
\begin{equation}
r=\sqrt{x^2+y^2+z^2}.
\end{equation}

Now we consider the extension of the exterior Schwarzschild spacetime in two
opposite directions.

We notice from the transformation (2) that when $\theta =\pi /2$ and $\phi
=0,\pi $ then 
\begin{equation}
y=z=0\qquad \qquad and\qquad \qquad x=\pm r\qquad \qquad (r>2M),
\end{equation}
where $x=+r$ corresponds to $\phi =0$ and $x=-r$ corresponds to $\phi =\pi $%
. That is, while $x>0$ corresponds to one direction, $x<0$ corresponds to
the opposite one. In this situation, from (3) and (4) one has 
\begin{equation}
ds^2=-(1-{\frac{2M}{\left| x\right| }})dt^2+{\frac{{\left| x\right| }}{{%
\left| x\right| }-2M}}dx^2\qquad \qquad \forall (t,x,y,z)\in \widetilde{x}%
_{+}\cup \widetilde{x}_{-},
\end{equation}
where $\widetilde{x}_{+}\equiv \{(t,x,y,z)|\infty >t>-\infty ,\infty
>x>2M,y=0,z=0\}$ is a portion of the exterior Schwarzschild spacetime in one
direction and $\widetilde{x}_{-}\equiv \{(t,x,y,z)|\infty >t>-\infty
,-2M>x>-\infty ,y=0,z=0\}$ is another portion of the spacetime in the
opposite direction. It can be verified that the transformation, 
\begin{equation}
\left\{ 
\begin{array}{c}
T=(\frac{{\left| x\right| }}{2M}-1)^{1/2}e^{{\left| x\right| /4M}}sh\frac
t{4M} \\ 
X=(\frac{{\left| x\right| }}{2M}-1)^{1/2}e^{{\left| x\right| /4M}}ch\frac
t{4M} \\ 
y=y \\ 
z=z
\end{array}
\right. \qquad \qquad \forall (t,x,y,z)\in \widetilde{x}_{+}
\end{equation}
and 
\begin{equation}
\left\{ 
\begin{array}{c}
T=(\frac{{\left| x\right| }}{2M}-1)^{1/2}e^{{\left| x\right| /4M}}sh\frac
t{4M} \\ 
X=-(\frac{{\left| x\right| }}{2M}-1)^{1/2}e^{{\left| x\right| /4M}}ch\frac
t{4M} \\ 
y=y \\ 
z=z
\end{array}
\right. \qquad \qquad \forall (t,x,y,z)\in \widetilde{x}_{-},
\end{equation}
will rewrite (6) in the form 
\begin{equation}
ds^2={\frac{32M^3e^{-{\left| x\right| /2M}}}{{\left| x\right| }}}%
(-dT^2+dX^2)\qquad \qquad \forall (T,X,y,z)\in \widetilde{X}_{+}\cup 
\widetilde{X}_{-},
\end{equation}
where $T$ and $X$ satisfy 
\begin{equation}
X^2-T^2=({\frac{{\left| x\right| }}{2M}}-1)e^{{\left| x\right| /2M}}\qquad
\qquad ({\left| x\right| }>2M),
\end{equation}
and $\widetilde{X}_{+}\equiv \{(T,X,y,z)|\infty >T>-\infty ,\infty
>X>0,y=0,z=0;X^2-T^2>0\}$, $\widetilde{X}_{-}\equiv \{(T,X,y,z)|\infty
>T>-\infty ,0>X>-\infty ,y=0,z=0;X^2-T^2>0\}$.

The above transformation, denoted $f_x$, is a 1-1 map,with $f_x(\widetilde{x}%
_{+})=\widetilde{X}_{+}$, $f_x(\widetilde{x}_{-})=\widetilde{X}_{-}$ and $%
f_x(\widetilde{x}_{+}\cup \widetilde{x}_{-})=\widetilde{X}_{+}\cup 
\widetilde{X}_{-}$. Note that, if relation (7) is applied in the
transformation for both $\widetilde{x}_{+}$ and $\widetilde{x}_{-}$, then
the map will no longer be 1-1 (even though the metric form of (9) is
maintained).

Keeping the metric form of (9) but allowing $X^2-T^2>-1$ instead of $%
X^2-T^2>0$ (these correspond to ${\left| x\right| }>0$ and ${\left| x\right| 
}>2M$ respectively), we have an extension of the part, $\widetilde{x}%
_{+}\cup \widetilde{x}_{-}$, of an exterior Schwarzschild spacetime,
confined in two opposite directions, in a manifold $\widetilde{X}$, $%
\widetilde{X}\equiv \{(T,X,y,z)|\infty >T>-\infty ,\infty >X>-\infty
,y=0,z=0\}$ which is an {\bf {\large R}}$^2$ space. The new extension is
2-dimensional and has the same $T-X$ spacetime diagram as has the
conventional Kruskal extension. (Note that, here ${\left| x\right| =r}$
according to (4) and (5).) However, for the new extension, while region $I$
of the diagram (i.e. $\widetilde{X}_{+}$) corresponds to a portion of the
exterior Schwarzschild spacetime in one direction, $\widetilde{x}_{+}$,
region $IV$ (i.e. $\widetilde{X}_{-}$) corresponds to none other than a
portion in the opposite direction, $\widetilde{x}_{-}$.

This result led us further to consider the extension of the exterior
Schwarzschild spacetime in an equatorial plane.

In the spacetime where 
\begin{equation}
\theta =\frac \pi 2,
\end{equation}
(1) becomes 
\begin{equation}
ds^2=-(1-{\frac{2M}r})dt^2+{\frac 1{1-{\frac{2M}r}}}dr^2+r^2d\phi ^2\qquad
\qquad (r>2M,\theta =\frac \pi 2).
\end{equation}
The following transformation 
\begin{equation}
\left\{ 
\begin{array}{c}
T=(\frac r{2M}-1)^{1/2}e^{r{/4M}}sh\frac t{4M} \\ 
R=(\frac r{2M}-1)^{1/2}e^{r{/4M}}ch\frac t{4M} \\ 
\theta =\theta \\ 
\phi =\phi
\end{array}
\right. \qquad \qquad \forall (t,r,\theta ,\phi )\in \widetilde{r}_{\pi /2},
\end{equation}
where $\widetilde{r}_{\pi /2}\equiv \{(t,r,\theta ,\phi )|\infty >t>-\infty
,\infty >r>2M,\theta =\frac \pi 2,2\pi \geq \phi \geq 0\}$, rewrites (12) in
the form 
\begin{equation}
ds^2={\frac{32M^3e^{-r{/2M}}}r}(-dT^2+dR^2)+r^2d\phi ^2\qquad \qquad \forall
(T,R,\theta ,\phi )\in \widetilde{R}_{\pi /2},
\end{equation}
where $T$, $R$ and $r$ are related by 
\begin{equation}
R^2-T^2=({\frac r{2M}}-1)e^{r/2M}\qquad \qquad (r>2M),
\end{equation}
and $\widetilde{R}_{\pi /2}\equiv \{(T,R,\theta ,\phi )|\infty >T>-\infty
,\infty >R>0,\theta =\frac \pi 2,2\pi \geq \phi \geq 0;R^2-T^2>0\}$.

The above transformation, denoted $f_r$, is a 1-1 map, with $f_r(\widetilde{r%
}_{\pi /2})=\widetilde{R}_{\pi /2}$, where $\widetilde{r}_{\pi /2}$ is an 
{\bf {\large R}}$^3$ space with an infinite length hollow column and $%
\widetilde{R}_{\pi /2}$ is an {\bf {\large R}}$^3$ with two head-to-head
opposite hollow cones extending to infinity.

In the same way, the extension of $\widetilde{R}_{\pi /2}$ is realized by
keeping the metric form of (14) but allowing $R^2-T^2>-1$ instead of $%
R^2-T^2>0$ (they correspond to $r>0$ and $r>2M$ respectively). In this
extension we find that a black hole is inside one of the two cones and a
white hole is inside the other cone. This extension is 3-dimensional and can
be realized by rotating the conventional Kruskal extension around its $T$
axis.

With the same method, the whole exterior Schwarzschild spacetime (in all
directions) can be extended by making the following transformation 
\begin{equation}
\left\{ 
\begin{array}{c}
T=(\frac r{2M}-1)^{1/2}e^{r{/4M}}sh\frac t{4M} \\ 
R=(\frac r{2M}-1)^{1/2}e^{r{/4M}}ch\frac t{4M} \\ 
\theta =\theta \\ 
\phi =\phi
\end{array}
\right. \qquad \qquad \forall (t,r,\theta ,\phi )\in \widetilde{r},
\end{equation}
where $\widetilde{r}\equiv \{(t,r,\theta ,\phi )|\infty >t>-\infty ,\infty
>r>2M,\pi \geq \theta \geq 0,2\pi \geq \phi \geq 0\}$, which rewrites (1) in
the form 
\begin{equation}
ds^2={\frac{32M^3e^{-r/2M}}r}(-dT^2+dR^2)+r^2(d\theta ^2+\sin ^2\theta d\phi
^2)\qquad \qquad \forall (T,R,\theta ,\phi )\in \widetilde{R},
\end{equation}
where 
\begin{equation}
R^2-T^2=({\frac r{2M}}-1)e^{r/2M}\qquad \qquad (r>2M),
\end{equation}
and $\widetilde{R}\equiv \{(T,R,\theta ,\phi )|\infty >T>-\infty ,\infty
>R>0,\pi \geq \theta \geq 0,2\pi \geq \phi \geq 0;R^2-T^2>0\}$. Then,
keeping the metric form of (17), we extend the range of $R^2-T^2>0$ to $%
R^2-T^2>-1$ (they correspond to $r>2M$ and $r>0$ respectively). This yields 
\begin{equation}
ds^2={\frac{32M^3e^{-r/2M}}r}(-dT^2+dR^2)+r^2(d\theta ^2+\sin ^2\theta d\phi
^2)\qquad \qquad \forall (T,R,\theta ,\phi )\in \widetilde{K},
\end{equation}
where 
\begin{equation}
R^2-T^2=({\frac r{2M}}-1)e^{r/2M}\qquad \qquad (r>0),
\end{equation}
and $\widetilde{K}\equiv \{(T,R,\theta ,\phi )|\infty >T>-\infty ,\infty
>R>0,\pi \geq \theta \geq 0,2\pi \geq \phi \geq 0;R^2-T^2>-1\}$.

This extension is 4-dimensional, and contains the extension of the
equatorial plane as a 3-dimensional section, and the conventional extension
as a 2-dimensional section. From the point of view of the 4-dimensional
extension, region $IV$ of the conventional extension does not belong to a
different universe but is a portion of the same exterior Schwarzschild
spacetime that contains region $I$. Particles can move from one to the
other. For example, the world line of a particle in a circular orbit outside
the Schwarzschild black hole (with $r=$ constant $>2M$ ) can be a line
around the $T$ axis and towards the direction of $T$ from region $I$ (where $%
\theta =\frac \pi 2,\phi =0$) through the regions of ($\theta =\frac \pi
2,0<\phi <\pi $) to region $IV$ (where $\theta =\frac \pi 2,\phi =\pi $),
shown in the spacetime diagram of the 3-dimensional head-to-head double-cone
extension. \\

\vspace{5mm}

\begin{flushleft}
{\bf {\large Acknowledgements}}
\end{flushleft}

It is my pleasure to thank Dr. Tao Kiang for helpful discussion. Thanks are
also given to Professors G. Z. Xie, Xue-Tang Zheng and Shi-Min Wu for their
help. This work was supported by the United Laboratory of Optical Astronomy,
CAS, the Natural Science Foundation of China, and the Natural Science
Foundation of Yunnan.

\vspace{35mm}

\begin{flushleft}
{\bf {\large References}}
\end{flushleft}

\begin{verse}
1. M. D. Kruskal, {\it Phys. Rev.} {\bf 119}, 1743 (1960).\\

2. R. M. Wald, {\it General Relativity} (The University of Chicago Press,
Chicago, 1984).\\
\end{verse}

\end{document}